\documentclass[aps,prl,twocolumn,showpacs,superscriptaddress,floatfix,nofootinbib]{revtex4}
\usepackage{graphicx}
\usepackage{amsmath}
\usepackage{epsfig}
\usepackage{helvet}
\usepackage{amssymb}

\newcommand{\be}{\begin{equation}}
\newcommand{\ee}{\end{equation}}
\newcommand{\bea}{\begin{eqnarray}}
\newcommand{\eea}{\end{eqnarray}}

\newcommand{\tr}{\mbox{tr}}
\newcommand{\bra}[1]{\mbox{$\langle #1 |$}}
\newcommand{\ket}[1]{\mbox{$| #1 \rangle$}}
\newcommand{\braket}[2]{\mbox{$\langle #1  | #2 \rangle$}}
\newcommand{\proj}[1]{\mbox{$|#1\rangle \!\langle #1 |$}}

\def\tr{ \mbox{tr}}

\begin{document}

\title{
A class of quantum many-body states that can be efficiently
simulated}

\author{G. Vidal}
\affiliation{School of Physical Sciences, the University of
Queensland, QLD 4072, Australia}
\date{\today}

\begin{abstract}
We introduce the multi-scale entanglement renormalization ansatz
(MERA), an efficient representation of certain quantum many-body
states on a D-dimensional lattice. Equivalent to a quantum circuit
with logarithmic depth and distinctive causal structure, the MERA
allows for an exact evaluation of local expectation values. It is
also the structure underlying entanglement renormalization, a
coarse-graining scheme for quantum systems on a lattice that is
focused on preserving entanglement.

\end{abstract}


\maketitle

A better understanding of quantum entanglement has enabled
significant progress in the numerical simulation of quantum
many-body systems over the last few years
\cite{vi1D,veci2D,tree,other1}. Building on the density matrix
renormalization group (DMRG) \cite{dmrg}--a well-established
technique for one-dimensional (1D) systems on a lattice--, new
insight from quantum information science has led, e.g., to efficient
algorithms to simulate time-evolution \cite{vi1D} and address 2D
systems \cite{veci2D}.

A key ingredient of such algorithms is the use of a network of
tensors to efficiently represent quantum many-body states. Examples
of tensor networks include matrix product states (MPS) \cite{mps}
for 1D systems, tree tensor networks (TTN) \cite{tree} for systems
with tree shape, and projected entangled-pair states (PEPS)
\cite{veci2D} for 2D systems and beyond, the three structures
differing in the graph that defines how the tensors are
interconnected into a network: the graphs for MPS, TTN and 2D PEPS
are, respectively, a chain, a tree and a 2D lattice. Importantly,
from these tensor networks the expectation value of local
observables can be computed efficiently. But whereas from a MPS and
a TTN such calculations are exact, from a PEPS --which has a much
wider range of applications-- local expectation values can only be
obtained efficiently after a number of approximations.

In this manuscript we present the {\em multi-scale entanglement
renormalization ansatz} (MERA), a structure that efficiently encodes
quantum many-body states of D-dimensional lattice systems and from
which local expectation values can be computed exactly. A MERA
consists of a network of {\em isometric} tensors in D+1 dimensions,
where the extra dimension can be interpreted in two alternative
ways: either as the time of a peculiar class of quantum
computations, or as parameterizing different length scales in the
system, according to successive applications of a lattice
coarse-graining procedure known as {\em entanglement
renormalization} \cite{viER}. Focused on preserving entanglement,
MERA is a promising candidate to describe emergent quantum
phenomena, including quantum phase transitions, quasi-particle
excitations and topological order. Here we establish its connection
to entanglement renormalization and explore some of its basic
properties: efficient contractibility of the network, leading to an
efficient evaluation of local expectation values; inherent support
of algebraically decaying correlations and of an area law for
entanglement; versatility to adapt to both the local and global
structure of the system's underlying lattice; and ability to
naturally assimilate symmetries such as invariance under
translations or rescaling -- which result in substantial gains in
computational efficiency.

\begin{figure}
  \includegraphics[width=8.5cm]{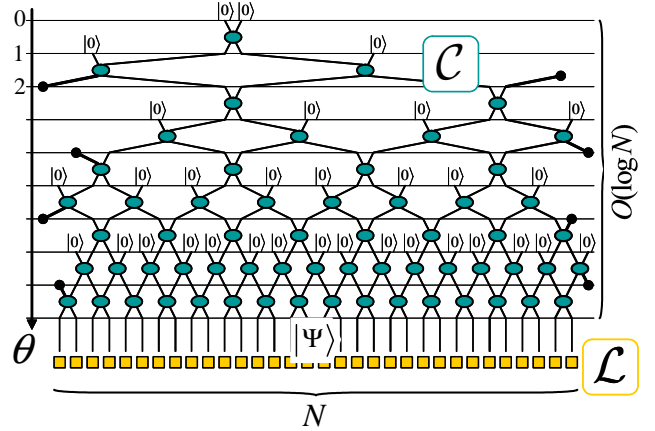}\\
\caption{Quantum circuit ${\cal C}$ transforms state
$\ket{0}^{\otimes N}$ into the $N$-site state $\ket{\Psi}$ of a 1D
lattice ${\cal L}$. ${\cal C}$ contains $2N-1$ gates organized in
$O(\log N)$ layers labeled by a discrete time $\theta$.}
\label{fig:circuit}
\end{figure}

Let us consider a square lattice $\cal{L}$ in $D$ spatial dimensions
consisting of $N$ sites, where each site $s\in {\cal L}$ is
described by a complex vector space $\mathbb{V}$ of finite dimension
$\chi$. Let $\ket{\Psi} \in \mathbb{V}^{\otimes N}$ denote a pure
state of lattice ${\cal L}$ and
$\rho^{[s]}=\tr_{\bar{s}}(\proj{\Psi})$ its reduced density matrix
for site $s$. We study states $\ket{\Psi}$ that can be generated by
means of a certain quantum circuit $\cal{C}$ of depth $\Theta \equiv
2\log_2(N)-1$, see Fig. (\ref{fig:circuit}) for the $D=1$ case. Each
site in $\cal{L}$ corresponds to one outgoing wire of $\cal{C}$, and
we label both with the same index $s$.

\begin{figure}
  \includegraphics[width=8.5cm]{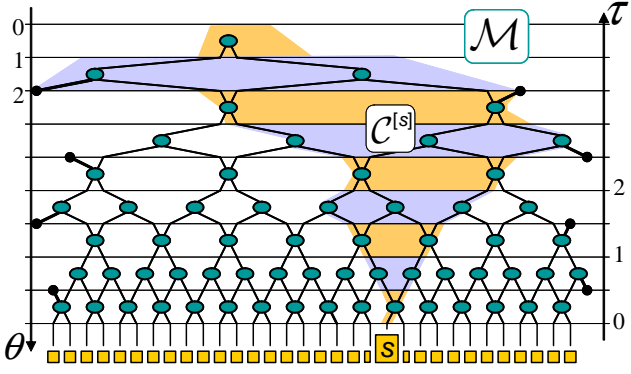}\\
\caption{A MERA ${\cal M}$ inherits the causal structure of quantum
circuit ${\cal C}$: the causal cone ${\cal C}^{[s]}$ for site $s$
has bounded width. When reversing the arrow of time $\theta$, ${\cal
M}$ implements entanglement renormalization
transformations.}\label{fig:MERA}
\end{figure}

\begin{figure}
  \includegraphics[width=8.5cm]{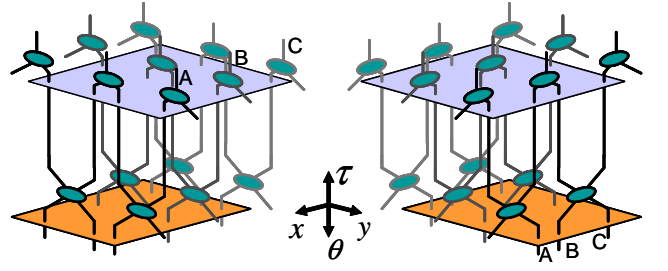}\\
\caption{Detail of a causal cone in a 2D MERA, implying that its
width is bounded. $3\times 3$ sites (bottom left) are mapped into
$3\times 3$ sites (top right) by layers of tensors that act along
the $y$ and $x$ direction [some stages involve $3\times 4$
sites].}\label{fig:2D}
\end{figure}

The {\em causal cone} ${\cal C}^{[s]}$ of outgoing wire $s$, defined
as the set of gates and wires that may influence state $\rho^{[s]}$,
plays an important role in our discussion. The key feature of
circuit ${\cal C}$ is that each causal cone ${\cal C}^{[s]}$ has
{\em bounded width}, that is, the number of wires in a time slice
${\cal C}^{[s]}_{\theta}$ of ${\cal C}^{[s]}$ is bounded by a
constant that is independent of $N$. [Here $\theta$ labels a
discrete time, with $\theta\in \{0,1,\cdots,\Theta\}$]. For
instance, the causal cone of any outgoing wire in Fig.
(\ref{fig:circuit}) involves at most four wires at any time
$\theta$, as highlighted in Fig. (\ref{fig:MERA}). More generally,
the causal cone of outgoing wires in the $D$-dimensional case can be
seen to comprise at most $3^{D-1}4$ wires at a given time $\theta$,
see Fig. (\ref{fig:2D}).

A MERA for $\ket{\Psi}$ is a tensor network ${\cal M}$ that
corresponds to quantum circuit ${\cal C}$ except for some minor
cosmetic changes. Each two-body unitary gate $u$ of ${\cal C}$ gives
rise to a tensor of ${\cal M}$. But incoming wires in state
$\ket{0}$ are eliminated in the tensor network, producing three
kinds of tensors: ($i$) the {\em top tensor} $t=u\ket{0}\ket{0}$ of
${\cal M}$ has two indices and is normalized to 1,
\begin{equation}\label{eq:top_tensor}
(t)_{\mu\nu} \equiv (u)_{\mu\nu}^{\alpha
\beta}|_{\alpha,\beta=0},~~~~
\sum_{\mu\nu}(t^{*})_{\mu\nu}(t)_{\mu\nu} = 1;
\end{equation}
($ii$) tensors in every second row are {\em isometries}
$w=u\ket{0}$,
\begin{equation}\label{eq:isometry}
(w)_{\mu\nu}^{\alpha} \equiv (u)_{\mu\nu}^{\alpha
\beta}|_{\beta=0},~~~~
\sum_{\mu\nu}(w^{*})_{\mu\nu}^{\alpha}(w)_{\mu\nu}^{\alpha'} =
\delta_{\alpha\alpha'};
\end{equation}
($iii$) the rest of tensors in ${\cal M}$ are unitary gates $u$,
\begin{eqnarray}\label{eq:unitary1}
\sum_{\mu\nu}(u^{*})_{\mu\nu}^{\alpha\beta}(u)_{\mu\nu}^{\alpha'\beta'}
= \delta_{\alpha\alpha'}\delta_{\beta\beta'},\\
\sum_{\alpha\beta}(u^{*})_{\mu\nu}^{\alpha\beta}(u)_{\mu'\nu'}^{\alpha\beta}
= \delta_{\mu\mu'}\delta_{\nu\nu'},\label{eq:unitary2}
\end{eqnarray}
that we call {\em disentanglers} for reasons that will become clear
later. Notice that the computational space required to store ${\cal
M}$ grows as $O(\chi^4N)$, that is, linearly in $N$, given that
there are $2N-1$ tensors and each tensor depends on at most $\chi^4$
parameters.

Thus a MERA is an {\em efficient} representation of $\ket{\Psi}$
consisting of a tensor network ${\cal M}$ in $D+1$ dimensions with
two properties: ($i$) tensors are constrained by Eqs.
(\ref{eq:top_tensor}), (\ref{eq:isometry}) or
(\ref{eq:unitary1})-(\ref{eq:unitary2}); ($ii$) each open wire $s$,
associated to one site of the underlying lattice ${\cal L}$, has a
causal cone ${\cal C}^{[s]}$ with bounded width. As a consequence of
this peculiar causal structure, the reduced density matrix of a
small number of lattice sites can be computed exactly with
remarkably small cost. In what follows $p_1$ and $p_2$ are integers
that depend on the spatial dimension $D$ of ${\cal L}$.

{\bf Lemma 1:} The one-site density matrix $\rho^{[s]}$ can be
computed from a MERA with time $O(\chi^{p_1} \log N)$.

\begin{figure}
  \includegraphics[width=8.5cm]{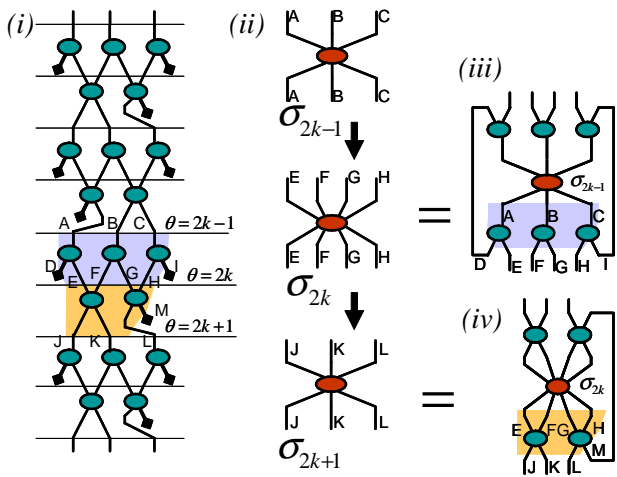}
\caption{Efficient computation of $\rho^{[s]}$. ($i$) Part of the
causal cone ${\cal C}^{[s]}$ of $\rho^{[s]}$, $D=1$. ($ii$) How to
compute $\sigma_{2k+1}$ from $\sigma_{2k-1}$ through simple tensor
multiplications ($iii$) and ($iv$).}\label{fig:efficient}
\end{figure}

{\bf Proof:} For each time slice ${\cal C}^{[s]}_{\theta}$ of the
causal cone ${\cal C}^{[s]}$, we compute its reduced density matrix
$\sigma_{\theta}$. As shown in fig. (\ref{fig:efficient}) for $D=1$,
$\sigma_{\theta+\!1}$ can be obtained from $\sigma_{\theta}$ with a
cost polynomial in $\chi$ and independent of $N$. Recall, finally,
that $\rho^{[s]} = \sigma_{\Theta}$ and that $\Theta = 2\log_2 (N)
-1$.

{\bf Lemma 2:} The two-site density matrix $\rho^{[s_1s_2]}$ can be
computed with $O(\chi^{p_2} \log N)$ time.

\begin{figure}
  \includegraphics[width=8.5cm,height=5cm]{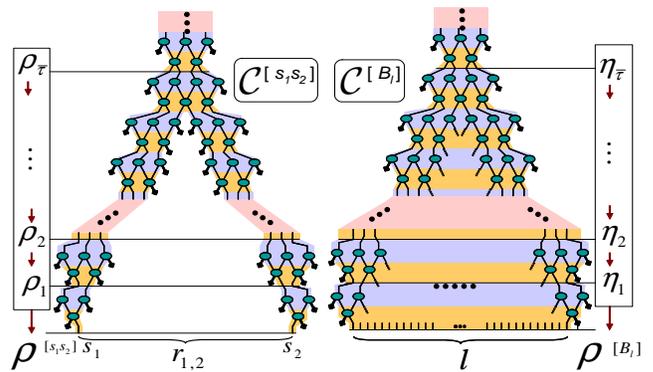}\\
\caption{Causal cones ${\cal C}^{[s_1s_2]}$ and ${\cal C}^{[B_l]}$
for a two-site density matrix $\rho^{[s_1 s_2]}$ (left) and an
$l$-site density matrix $\rho^{[B_l]}$ (right).}\label{fig:rho2}
\end{figure}

{\bf Proof:} Again, the causal cone ${\cal C}^{[s_1s_2]}$ has
logarithmic depth and a width independent of $N$, see Fig.
(\ref{fig:rho2}), and the reduced density matrix $\sigma_{\theta+1}$
for time slice ${\cal C}^{[s_1s_2]}_{\theta+\!1}$ can be obtained
from $\sigma_{\theta}$ for ${\cal C}^{[s_1s_2]}_{\theta}$ at bounded
cost.

The above lemmas can be easily extended to a $k$-site reduced
density matrix $\rho^{[s_1\cdots s_k]}$ (with computational time
scaling exponentially in $k$) and they imply that the expectation
value of local observables, such as two-site correlators
\begin{equation}\label{eq:2site_corr}
    C_2(s_1,s_2)  \equiv
    \bra{\Psi}A^{[s_1]}B^{[s_2]}\ket{\Psi} = \tr
    (\rho^{[s_1s_2]}A^{[s_1]}B^{[s_2]}),
\end{equation}
can be computed efficiently.

We have defined a MERA for $\ket{\Psi}$ in terms of a quantum
circuit ${\cal C}$ that transforms a product state $\ket{0}^{\otimes
N}$ into $\ket{\Psi}$ by means of $\Theta$ layers of unitary gates.
Let $\ket{\Psi_{\tau}}$ denote [the non-trivial part of] the state
of circuit ${\cal C}$ at time $\theta=\Theta-2\tau$, with
$\ket{\Psi_0}\equiv \ket{\Psi}$. An alternative interpretation of
the MERA can be obtained by considering the sequence of states
$\{\ket{\Psi_0}, \ket{\Psi_1},\ket{\Psi_2}, \dots\}$, which
correspond to undoing the quantum evolution of ${\cal C}$ back in
time. Notice that $\ket{\Psi_{\tau+1}}$ is obtained from
$\ket{\Psi_{\tau}}$ by applying two layers of tensors in ${\cal M}$.
The first layer is made of disentanglers that transform
$\ket{\Psi_{\tau}}$ into a {\em less entangled} state
$\ket{\Psi_{\tau}'}$. The second layer is made of isometries that
combine pairs of nearest neighbor wires into single wires, turning
the state $\ket{\Psi_{\tau}'}$ of $N_{\tau}$ wires into the state
$\ket{\Psi_{\tau+1}}$ of $N_{\tau}/2$ wires, where $N_{\tau} =
2^{\Theta-\tau}$. That is, ${\cal M}$ implements a class of real
space coarse-graining transformations known as {\em entanglement
renormalization} \cite{viER}.

More generally, we can use the MERA to transform the sites of
lattice ${\cal L}_0 \equiv {\cal L}$, as well as operators defined
on ${\cal L}$ such as a Hamiltonian $H_0$, so as to obtain a
sequence of increasingly coarse-grained lattices $\{ {\cal L}_0,
{\cal L}_1, {\cal L}_2, \dots\}$ and corresponding effective
Hamiltonians $\{H_0, H_1, H_2, \dots\}$. Since an operator defined
on site $s\in {\cal L}$ is mapped into an operator contained in the
causal cone ${\cal C}^{[s]}$, {\em local} operators in ${\cal L}$
remain {\em local} when mapped into ${\cal L}_{\tau}$, $1\leq \tau
\leq \log_2(N)-1$. Notice that one site in ${\cal L}_{\tau}$ is
obtained from a D-dimensional hypercube with $O(2^{\tau})$ sites in
${\cal L}$. Thus $\ket{\Psi_{\tau}}$ can be understood as retaining
the structure of $\ket{\Psi}$ at length scales greater than
$2^{\tau/D}$.

In \cite{viER} it has been shown that a MERA can encode, in a
markedly more efficient way than a MPS \cite{mps_mera}, accurate
approximations to the ground state of a quantum critical system in a
1D lattice. There correlators $C_2(s_1,s_2)$ decay as a power law
with the distance $r$ between sites $s_1$ and $s_2$ and the
entanglement entropy $S(\rho^{[B_l]})$ of a block $B_l$ of $l$
adjacent sites scales as $\log l$ \cite{viscal}. Next we give an
intuitive justification why MERA supports algebraic decay of
correlations in any dimension $D$. We also show that a MERA can
carry an amount of block entanglement compatible with logarithmic
scaling in 1D and with a boundary law $l^{D-1}$ for $D>1$.

It follows from the $\lambda$-shaped causal cone ${\cal
C}^{[s_1s_2]}$ that we can compute $\rho^{[s_1s_2]}$ from the
density matrix $\rho_{\bar{\tau}}$ corresponding to a hypercube of
(at most) $4^D$ sites of ${\cal L}_{\bar{\tau}}$, where $\bar{\tau}
\approx D\log_2 r$, by means of a sequence of density matrices
$\{\rho_{\bar{\tau}}, \dots, \rho_2, \rho_1, \rho^{[s_1s_2]}\}$, see
Fig. (\ref{fig:rho2}). That is, $\rho^{[s_1s_2]}$ is obtained from
$\rho_{\bar{\tau}}$ after $O(\log r)$ transformations. If, as it can
be argued in a (scale invariant) critical ground state, each of
these transformations reduces correlations by a constant factor
$z<1$, we readily obtain a power law scaling for $C_2(s_1,s_2)$,
\begin{equation}\label{eq:correlations}
    C_2(s_1,s_2) \approx z^{\log r} = r^{-q}, ~~~~~~~ q = \log
    \frac{1}{z}.
\end{equation}
This property is enabled by the fact that two sites at a distance
$r$ in ${\cal L}$ are connected through a path of length $O(\log r)$
in ${\cal M}$, and it indicates that a MERA is particularly suited
to describe states with quasi-long-range order, such as critical
ground states.

We now consider the entropy $S(\rho^{[B_l]})$ for the density matrix
$\rho^{[B_l]}$ of a hypercube $B_l$ made of $l^D$ sites. The causal
cone of $B_l$ shrinks exponentially fast with $\tau$ and we see
that, once more, we can compute $\rho^{[B_l]}$ from the density
matrix $\eta_{\bar{\tau}}$ for a hypercube made of (at most) $4^D$
sites of lattice ${\cal L}_{\bar{\tau}}$, for $\bar{\tau} \approx
D\log_2 l$, through a sequence of density matrices
$\{\eta_{\bar{\tau}}, \cdots, \eta_{1}, \rho^{[B_l]}\}$, see Fig.
(\ref{fig:rho2}). Density matrix $\eta_{\tau}$ is obtained from
$\eta_{\tau+1}$ by: ($i$) applying a layer of isometries and a layer
of disentanglers that do not change the entropy of $\eta_{\tau+1}$;
($ii$) tracing out $n_{\tau} = O(2^{(\bar{\tau}-\tau)(D-1)/D})$
boundary sites. Tracing out one boundary site increases the entropy
by at most $\log_2\chi$ bits, the total increase in entropy $\Delta
S_{\tau}$ being at most $n_{\tau}\log_2(\chi)$ bits. Thus the
entropy of $\rho^{[B_l]}$ fulfils
\begin{equation}\label{eq:entropyD1}
    S(\rho^{[B_l]}) - S(\eta_{\bar{\tau}}) = \sum_{\tau = 1}^{\bar{\tau}}
    \Delta S_{\tau} \leq \log_2(\chi)\sum_{\tau = 1}^{\bar{\tau}}
    n_{\tau},
\end{equation}
where $S(\eta_{\bar{\tau}})$ is at most $4^{D}\log_2\chi$.

In a $1D$ MERA the number $n_{\tau}$ of sites that are traced out is
bounded by a constant $c$ [i.e., a hypercube has only two boundary
sites] for any $\tau$, $1 \leq \tau \leq \bar{\tau} = O(\log l)$,
and
\begin{equation}\label{eq:entropyD1}
    S(\rho_{\{l\}}) - S(\eta_{\bar{\tau}}) \leq \log_2(\chi) c\bar{\tau}= O(\log l).
\end{equation}
As numerically confirmed in [viEr,evER], in a MERA for critical 1D
ground states $\Delta S_{\tau}$ is independent of $\tau$, leading to
the logarithmic scaling $S(\rho^{[B_l]}) \approx \log (l)$, whereas
for a non-critical ground state $\Delta S_{\tau}$ vanishes for $\tau
\gg \log_2\xi$, where $\xi$ is the correlation length, so that
$S(\rho^{[B_l]})$ saturates for $l \gg \xi$.

In $D>1$, instead, the $n_{\tau}$ decays exponentially with $\tau$,
the upper bound for the $S(\rho^{[B_l]})$ being dominated by the
contribution from small $\tau$,
\begin{eqnarray}\label{eq:entropyD}
S(\rho^{[B_l]}) - S(\eta_{\bar{\tau}}) &\leq&
\log_2(\chi)\sum_{\tau = 1}^{\bar{\tau}} 2^{(\bar{\tau}-\tau) \frac{D-1}{D}} \nonumber\\
\approx ~\log_2(\chi)~ 2^{\bar{\tau}\frac{D-1}{D}} &\approx&
\log_2(\chi)~
2^{D\log_2 (l) \frac{D-1}{D}} \nonumber\\
&=& \log_2(\chi)~l^{D-1}.
\end{eqnarray}
That is, a MERA in $D>1$ supports block entanglement that scales at
most according to a boundary law $S^{[B_l]} \approx l^{D-1}$.

For the sake of concreteness, we have analyzed the case where ${\cal
L}$ is a square lattice. The structure of the MERA, however, can be
adapted to a more generic lattice, with arbitrary local, geometric
and topological properties, while preserving its distinctive causal
structure. For instance, in $D=2$ dimensions a specific MERA can be
built to represent states of a triangular lattice; or of a lattice
with random vacancies or linear or bulk defects; or to account for a
variety of boundary conditions (e.g., plane, cylinder, sphere or
torus). In addition, the number of levels $\chi$ can vary throughout
${\cal M}$. Adjusting the MERA to the specifics of a problem often
leads to computational gains.

In particular, the symmetries of $\ket{\Psi}$ can be assimilated
into the MERA. An internal symmetry, such as $SU(2)$ invariance,
results in a series of constraints for the tensor in ${\cal M}$
\cite{suER}, which then depend on less parameters. More substantial
gains are obtained when $\ket{\Psi}$ is invariant under translations
[that is, cyclic shifts by one lattice site in a system with
periodic boundary conditions], since all the tensors in a layer of
${\cal M}$ can be chosen to be the same and the MERA depends only on
about $\chi^4\log_2 N$ parameters. But the most dramatic savings
occur for states that are invariant under entanglement
renormalization transformations, even in an infinite lattice
\cite{viER,evER}. Here all tensors in ${\cal M}$ are the same and
the MERA depends on just $O(\chi^4)$ parameters. Scale invariant
critical ground states \cite{evER} can be shown to belong to this
class.

We conclude with a few pointers to future work. On the one hand,
most techniques to simulate quantum systems with a MPS can be
generalized to a MERA. This include algorithms to compute the ground
and thermal states, and to simulate time evolution. For instance, in
order to update ${\cal M}$ after the action of a unitary gate
$V^{[s_1s_2]}$ on sites $s_1$ and $s_2$, only the tensors in the
causal cone ${\cal C}^{[s_1s_2]}$ need to be modified, with a cost
logarithmic in $N$.

On the other hand, the potential of a MERA is not restricted to the
representation of individual states. Notice that by feeding the
incoming wires, labeled by $r$, of quantum circuit $\cal C$ with
state $\otimes_{r=1}^N \ket{\phi_r^{[r]}}$ instead of
$\otimes_{r=1}^N\ket{0^{[r]}}$, we can generate an infinite family
of entangled states $\ket{\Psi_{\{\phi_r\}}}$, all represented by a
MERA that only differs in the isometric tensors and such that, given
the unitarity of ${\cal C}$, fulfill
\begin{equation}\label{eq:scalar_product}
\braket{\Psi_{\{\phi_r\}}}{\Psi_{\{\phi'_r\}}} = \prod_r
\braket{\phi^{[r]}_r}{\phi'^{[r]}_r}.
\end{equation}
This can be used to encode, in just one (generalized) MERA, not only
the ground state of a lattice system but also its quasi-particle
excitations. This is also used, in systems with topological quantum
order, to store topological information in the top tensor of a MERA
\cite{Topo}.

The author acknowledges support from the Australian Research Council
through a Federation Fellowship.


\begin{thebibliography}{99}
\bibitem{mps_mera} In an MPS, correlations between left and right halves of a
chain are accounted for through $\chi_{M\!P\!S}$ terms in the
Schmidt decomposition. A MERA can account for about $2\chi\log_2 N$
terms in the Schmidt decomposition. Thus, it takes an MPS with
$\chi_{MPS} \approx 2\chi\log_2 N$ to represent a state stored with
a MERA of $\chi$-level wires.
\end{thebibliography}
\end{document}